\input{aipcheck}

\documentclass[
    ,final            % use final for the camera ready runs
  ]
  {aipproc}

\layoutstyle{6x9}

\usepackage{graphicx}

\def\De{\Delta}

\def\anu{\alpha_\nu}
\def\anub{\alpha_{\bar{\nu}}}

\def\to{\rightarrow}

\def\beq{\begin{eqnarray}}
\def\eeq{\end{eqnarray}}

\def\numu{\nu_\mu}
\def\numub{\bar{\nu}_\mu}
\def\nub{\bar{\nu}}

\def\mum{\mu^{-}}

\def\pip{\pi^{+}}
\def\pim{\pi^{-}}
\def\thetmu{\theta_{\mu}}

\begin{document}

\title{New Results from MiniBooNE Charged-Current Quasi-Elastic Anti-Neutrino Data}


\author{Joseph Grange, for the MiniBooNE collaboration}{
  address={jgrange@phys.ufl.edu}
}

\begin{abstract}

MiniBooNE anti-neutrino charged-current quasi-elastic (CCQE) data is compared to model predictions.  The main background of neutrino-induced events is examined first, where three independent techniques are employed.  Results indicate the neutrino flux is consistent with a uniform reduction of $\sim$ 20\% relative to the largely uncertain prediction. After background subtraction, the $Q^{2}$ shape of $\numub$ CCQE events is consistent with the model parameter $M_{A}$ = 1.35 GeV determined from MiniBooNE $\numu$ CCQE data, while the normalization is $\sim$ 20\% high compared to the same prediction.

\end{abstract}

\maketitle

\section{Introduction}

The charged-current quasi-elastic interaction (CCQE, $\nu_{l} + N \to l + N'$) has become a hot topic lately, as many recent measurements on nuclear targets are in tension with previous results mostly from light nuclear targets.  Even more recently, theoretical work looking to explain these data suggest the widely-used Fermi Gas model~\cite{RFG} is insufficient for scattering in a nuclear environment, where intra-nuclear correlations may play a larger role than previously expected~\cite{Martini2,Amaro,Nieves2}.  Significant effects from multi-nucleon correlations in a nuclear environment are also seen in electron scattering data~\cite{Carlson}.

Anti-neutrino CCQE provides a strong test of the underlying physics by probing a different mix of axial and vector terms compared to neutrino-induced CCQE.  No anti-neutrino CCQE measurement has been made below 1~GeV, and few exist on nuclear targets.  The work presented here represents one of the final steps of MiniBooNE data analysis before producing absolute, differential and double-differential anti-neutrino CCQE cross sections on a nuclear target around 1~GeV.

%At the very least, compared to neutrinos the effect observed in electron scattering data appears differently in the anti-neutrino cross section due to the axial-vector interference term.  No anti-neutrino induced CCQE measurement exists below 1~GeV, and very few exist on nuclear targets.  The work presented here represents one of the final steps of MiniBooNE data analysis before producing absolute, differential and double-differential anti-neutrino CCQE cross sections.

%Both flux and cross section effects serve to amplify the neutrino contribution to the anti-neutrino mode sample compared to the amount of anti-neutrino interactions in the neutrino-mode data.  

The MiniBooNE detector is non-magnetized, so a dedicated study of the $\numu$ background is necessary before a meaningful appraisal of the anti-neutrino CCQE rate and shape can be made.  Three independent techniques to measure the neutrino component of the anti-neutrino mode data are presented, followed by comparisons between anti-neutrino CCQE data and model predictions.

\section{Measurements of the Neutrino Flux in Anti-Neutrino Running}
\label{wsMeas}

\begin{figure}[h]
$\begin{array}{cc}
\includegraphics[scale=0.38]{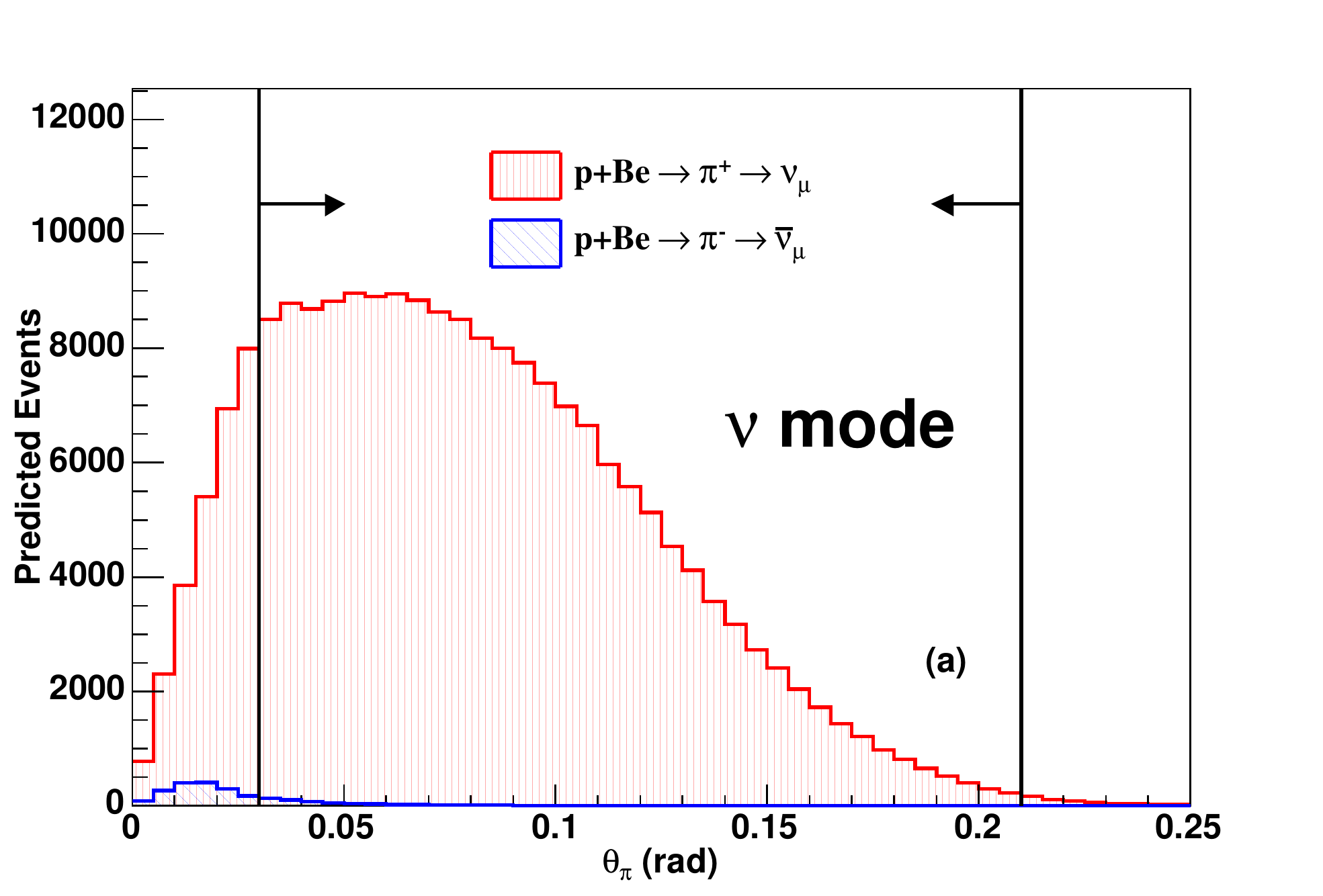} & 
\includegraphics[scale=0.38]{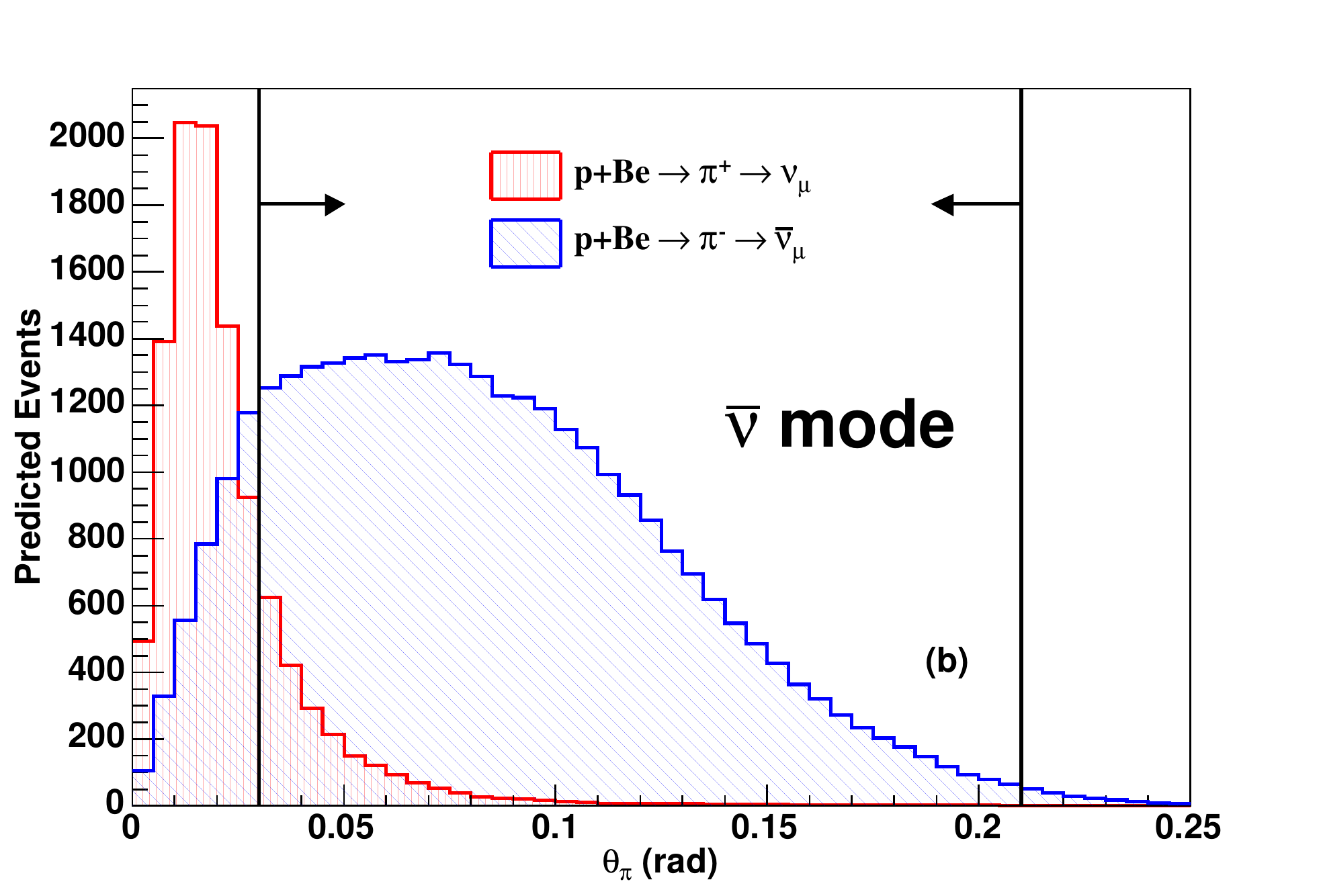} \\
\end{array}$
\caption{(Color online) Predicted angular distributions of pions at the proton target with respect to the incident beam ($\theta_{\pi}$) producing $\numu$ and $\numub$ in (a) neutrino mode and (b) anti-neutrino mode.  Only pions leading to events in the detector are shown, and arrows indicate the region where HARP data~\cite{HARP} are available.  Both cross section and flux effects conspire to suppress the wrong-sign contribution to neutrino-mode data, while their presence is amplified in anti-neutrino mode.}
\label{fig:prodSpace}
\end{figure}

The MiniBooNE flux prediction uses dedicated $\pi^\pm$ double-differential cross section measurements from the HARP experiment~\cite{HARP}.  As Figure~\ref{fig:prodSpace} shows, most of the production phase space of so-called ``right-sign'' particles (i.e., neutrinos in the neutrino-mode beam and anti-neutrinos in the anti-neutrino mode beam) are covered by the HARP data.  However, ``wrong-sign'' particles mostly arise from very forward-going pions that escape deflection by the magnetic horn, and this region is not covered by the HARP data.  Most of the wrong-sign MiniBooNE flux prediction is therefore extrapolated from the existing HARP data and is largely uncertain.  

%However, for the wrong-sign parent pions that generally escape deflection via low-angle and high-energy production, only a small subset of their production region is covered by HARP data.  

%\begin{figure}
%\begin{tabular}{cc}
%\includegraphics[scale=0.38]{./piAngleb_SB2.pdf} & 
%\includegraphics[scale=0.38]{./piAnglea_SB2.pdf}
% \end{tabular}
%\caption{(Color online) Predicted angular distributions of pions at the proton target with respect to the incident beam ($\theta_{\pi}$) producing $\numu$ and $\numub$ in (a) neutrino mode and (b) anti-neutrino mode.  Only pions leading to events in the detector are shown, and arrows indicate the region where HARP data~\cite{HARP} are available.  Both cross section and flux effects conspire to suppress the wrong-sign contribution to neutrino-mode data, while their presence is amplified in anti-neutrino mode.}
%\label{fig:prodSpace} 
%\end{figure}	

Clearly, a dedicated study of the neutrino content of the primarily anti-neutrino beam must be made before precise anti-neutrino CCQE cross sections can be made.  Three independent and complementary measurements of this crucial background are presented in this section: we evaluate charged-curent single $\pip$ (CC1$\pip$, $\nu_{l} + N \rightarrow l + N + \pi$) event rates, exploit $\mum$ capture in the detector medium, and finally the muon angular distribution of the CCQE sample is fit to a sum of the simulated $\numu$ and $\numub$ distributions to measure the neutrino component of the anti-neutrino beam.  In all three measurements, the relevant $\numu$ channels in simulation have been corrected to reflect the observed cross sections from neutrino-mode data, and therefore the measurements are interpreted as a flux calibration.  The CC1$\pip$ and CCQE angular fit methods and results are presented in more detail in Ref.~\cite{wsPaper}.  This work represents the first measurement of the neutrino component of an anti-neutrino beam using a non-magnetized detector.

%Three independent and complementary measurements of this crucial background are presented in this section: we evaluate charged-curent single $\pip$ (CC1$\pip$) event rates, exploit $\mum$ capture in the detector medium, and finally the $\mum$ and $\mup$ CCQE angular distributions are fit to data to measure the neutrino component of the anti-neutrino beam.

\subsection{CC1$\pip$ $\numu$ Flux Measurement}
\label{sbsec:CC1pi}

By simple charge and lepton number conservation, the CC1$\pi$ interaction yields $\pip$ ($\pim$) for an incident $\nu$ ($\nub$).  In a carbon nuclear environment, stopped $\pim$ experience capture at a rate of $\sim$100\%~\cite{pimCap}.  In this analysis, pions are identified only by the decay chain leading to the production of electrons from decay-at-rest muons (``Michel'' electrons), so the selection requirement of two delayed Michel electrons (one each from the $\pi$ decay chain and the primary muon decay) in addition to the prompt muon largely rejects anti-neutrinos. Absolute and differential cross sections for CC1$\pip$ have been measured previously using the MiniBooNE neutrino-mode data~\cite{CCpip}, and these are applied to simulation in the present analysis.  In this way, the residual level of agreement between data and simulation measures the accuracy of the extrapolated flux prediction discussed in the previous section.  To check the shape of the flux prediction, the CC1$\pip$ sample is binned in reconstructed neutrino energy $E_{\nu}^{\mathrm{\De}}$ assuming $\De$(1232) resonance and a stationary proton target: 

\begin{equation}
\label{eqn:EnuPi}
E_{\nu}^{\mathrm{\De}}= \frac{2 \left(M - E_{B}\right) E_{\mu} - \left( E^{2}_{B} - 2 M E_{B} + m^{2}_{\mu} + \De M^{'2} \right)}{2\left[ \left(M - E_{B} \right) - E_{\mu} + p_{\mu}\,\textrm{cos}\,\theta_{\mu}\right]},
\end{equation}
  
\noindent where $E_{B} = 34$~MeV is the binding energy, $m_{\mu}$ is the muon mass, $\De M^{'2} = M_{p}^{2} - M_{\De}^{2}$, where $M_{\De}$ ($M$) is the $\De$(1232) (nucleon) mass, $p_{\mu}$ is the muon momentum, and $\thetmu$ is the outgoing muon angle relative to the incoming neutrino beam.  Results summarized in Figure~\ref{fig:wsSum} indicate the $\numu$ flux prediction is overestimated by $\sim$ 24\% in normalization, while based on consistent results across bins of neutrino energy, the flux shape appears well-modeled.  This flux normalization (0.76 $\pm$ 0.11) is used exclusively in the subtraction of the $\numu$ contribution to the $\numub$ CCQE sample presented later, as this CC1$\pip$ analysis is the least model-dependent, carries the smallest uncertainty of the measurements and is entirely independent of the CCQE model.

%\begin{table}
%\begin{tabular}{lrrrrrrrrr}
%\hline
%\tablehead{1}{r}{b}{E$_{\nu}^{\mathrm{\De}}$ Range} & &
%\tablehead{1}{r}{b}{Data events} & &
%\tablehead{1}{r}{b}{Expected $\numu$} & &
%\tablehead{1}{r}{b}{Expected $\numub$} & & 
%\tablehead{1}{r}{b}{$\numu$ Flux Scale $\anu$}   \\
%\hline
%600 - 700 & & 465 & & 556 & & 104 & & 0.65 $\pm$ 0.10 \\
%700 - 800 & & 643 & & 666 & & 118 & & 0.79 $\pm$ 0.10 \\
%800 - 900 & & 573 & & 586 & & 97 & & 0.81 $\pm$ 0.10 \\
%900 - 1000 & &Ê495 & & 474 & & 78 & & 0.88 $\pm$ 0.11 \\
%1000 - 1200 & & 571 & & 646 & & 92 & & 0.74 $\pm$ 0.10 \\
%1200 - 2400 & & 521 & & 614 & & 74 & & 0.73 $\pm$ 0.15 \\
%\hline
%Inclusive & & 3268 & & 3542 & & 563 & & 0.76 $\pm$ 0.11 \\
%\hline
%\end{tabular}
%\caption{CC$1\pip$ sample details and $\numu$ flux component measurement.  The $\numu$ flux scale is found by calculating (observed events - expected $\numub$ events) / expected $\numu$ events.}
%\label{tbl:wsCCpi}
%\end{table}

%The prediction for the anti-neutrino contribution to the sample (primarily from decay-in-flight $\pim$) depends on the Rein-Sehgal~\cite{R-S} model for pion production, so a conservative 30\% error is assigned to the anti-neutrino contribution Because the selection between CC1$\pip$ cross section was measured in the same detector as the present This provides a powerful, model-independent check 

\subsection{$\mum$ Capture $\numu$ Flux Measurement}
\label{sbsec:mumCap}

Stopped $\mum$ undergo nuclear capture on carbon atoms at a rate of $\sim$ 8\%~\cite{mumCap}, so neutrino-induced CCQE events have a significantly lower probability for producing Michel electrons compared to anti-neutrino CCQE events.  For this analysis, two CCQE samples are formed.  The selection of the two samples is identical with the exception of a requirement of a single delayed Michel electron (hereafter referred to as the ``$\mu+e$'' sample) or no decay electrons (the ``$\mu$-only'' sample).  Linear scales applied to the $\numu$ and $\numub$ contributions ($\anu$ and $\anub$, respectively) in the simulation are determined by simultaneously requiring the ratio ($\mu$-only)/($\mu + e$) and the normalization to agree between data and the simulation.  This yields

\begin{eqnarray}
\label{eqn:solveIt}
\anu = \frac{(\mu\textrm{-only})^{\,\textrm{data}}\,(\mu+e)_{\numub}^\textrm{MC}-(\mu+e)^{\textrm{data}}\,(\mu\textrm{-only})_{\numub}^{\textrm{MC}}}{(\mu+e)_{\numub}^{\textrm{MC}}\,(\mu\textrm{-only})_{\numu}^{\textrm{MC}}-(\mu\textrm{-only})_{\numub}^{\textrm{MC}}\,(\mu+e)_{\numu}^{\textrm{MC}}} = 0.83 \pm 0.15 \\
\anub =\frac{(\mu\textrm{-only})^{\textrm{data}}\,(\mu+e)_{\numu}^{\textrm{MC}}-(\mu+e)^{\textrm{data}}\,(\mu\textrm{-only})_{\numu}^{\textrm{MC}}}{(\mu+e)_{\numu}^{\textrm{MC}}\,(\mu\textrm{-only})_{\numub}^{\textrm{MC}}-(\mu\textrm{-only})_{\numu}^{\textrm{MC}}\,(\mu+e)_{\numub}^{\textrm{MC}}} = 1.12 \pm 0.24, \end{eqnarray}

\noindent where, for example, $(\mu\textrm{-only})^{\textrm{data}}$ is the number of data events for the $\mu$-only sample and $(\mu+e)_{\numu}^{\textrm{MC}}$ is the predicted $\numu$ contribution to the $(\mu+e)$ sample.

The $\mu$-only and $\mu + e$ samples are dominated by the CCQE process, and simulation for both $\numu$ and $\numub$ CCQE have been modified to reflect the model parameters observed in neutrino-mode data~\cite{qePRD}.  The $\numu$ flux measurement, 0.83 $\pm$ 0.15, is consistent and complimentary with that found in the previous section.  The error on the $\anub$ rate scale is larger due to conservative uncertainties placed on the interaction to cover the extrapolated model parameter assumptions; fortunately, the $\numu$ flux scale $\anu$ is not very sensitive to the $\numub$ contributions.  Due to the limited statistics of the $\mu$-only sample, measurements are not made in bins of neutrino energy in order to check the $\numu$ flux shape as is done in the other methods.  

\subsection{CCQE Angular Fit $\numu$ Flux Measurement}
\label{sbsec:uzFits}

\begin{figure}[h]
\includegraphics[scale=0.46]{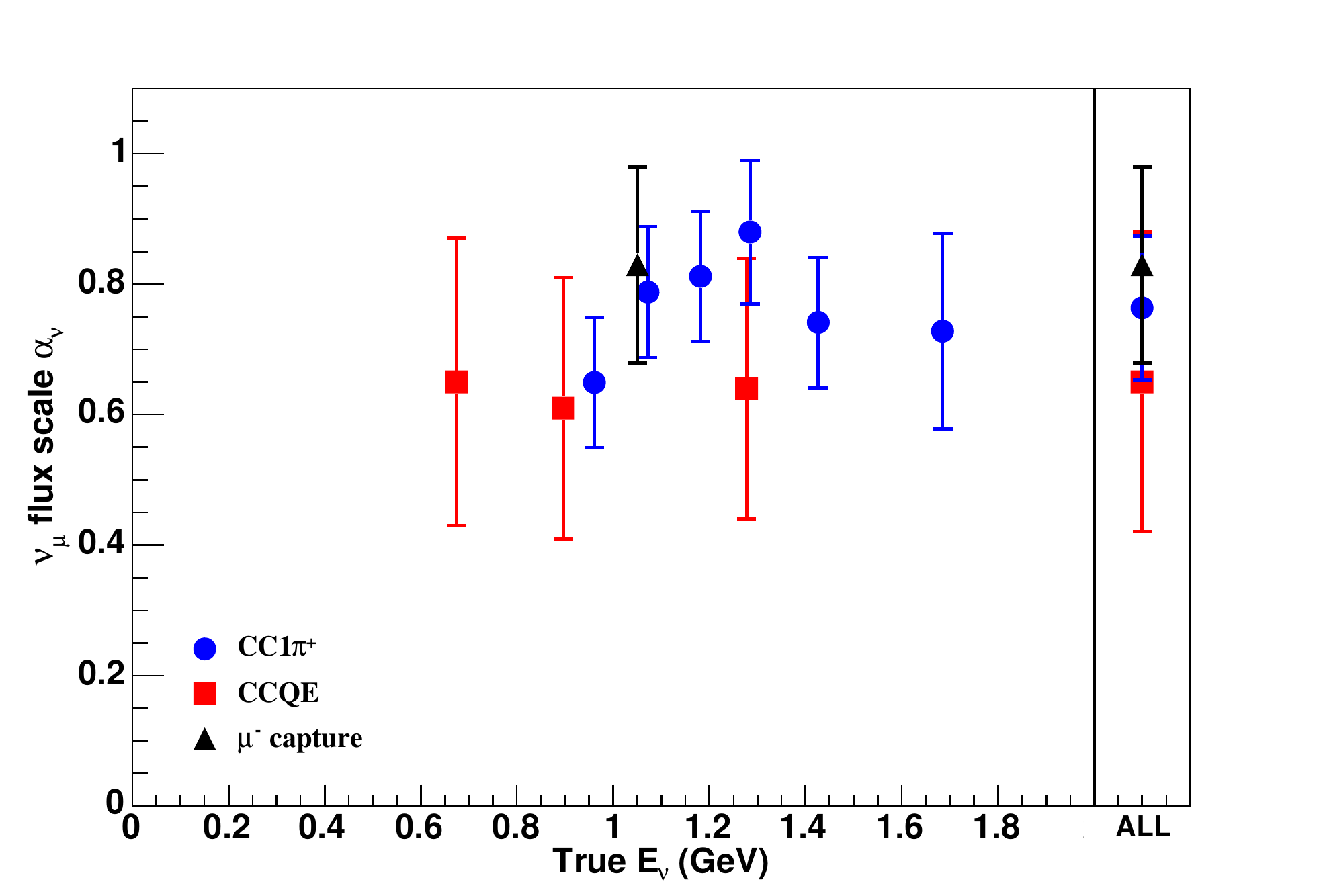}
\caption{(Color online) Summary of the three $\numu$ flux measurements in the anti-neutrino beam.  Measurements are placed at the mean of the generated energy distribution for each reconstructed energy sample.}
\label{fig:wsSum}
\end{figure}

Neutrino and anti-neutrino CCQE cross sections differ exclusively by an axial-vector interference term.  In particular, the contribution from $\numub$ is suppressed in the backward scattering region relative to the incoming neutrinos.  The $\mu+e$ sample used in the previous section is also employed in this analysis, but with slightly different selection cuts to enhance the CCQE purity.  Simulation is separated into the predicted $\numu$ and $\numub$ contributions, and a linear combination of these templates are fit to data.  Results binned in energy are presented in Figure~\ref{fig:wsSum}.  Appropriate to the dominant $\numub$ CCQE contribution to the sample, the reconstructed energy is found as in Eqn.~\ref{eqn:EnuPi} with the neutron mass substituted for the $\De$(1232) mass. 

%For consistency, the linear scale measured for $\numu$ ($\numub$) is labelled $\anu$ ($\anub$).

The angular fit finds results consistent with the CC1$\pip$ and $\mum$ capture measurements, namely that the $\numu$ flux is overestimated in normalization while the shape of the flux prediction is robust.

%\begin{table}
%\begin{tabular}{lrrrrrr}
%\hline
%\tablehead{1}{r}{b}{E$_{\nu}^{QE}$ Range} & &
%\tablehead{1}{r}{b}{$\anu$} & & 
%\tablehead{1}{r}{b}{$\anub$} \\  
%\hline
%0 - 600 & & 0.65 $\pm$ 0.22 & & 0.98 $\pm$ 0.18 \\
%600 - 900 & & 0.61 $\pm$ 0.20 & & 1.05 $\pm$ 0.19 \\
%\textgreater\, 900 & & 0.64 $\pm$ 0.20 & & 1.18 $\pm$ 0.21 \\
%\hline
%Inclusive & & 0.65 $\pm$ 0.23 & & 1.00 $\pm$ 0.22 \\
%\hline
%\end{tabular}
%\caption{CCQE angular fit results.  The results are consistent with an over-prediction of the $\numu$ contamination of anti-neutrino data.}
%\label{tbl:uzFitz}
%\end{table}

The results from all three analyses are consistent and complementary.  As shown in Figure~\ref{fig:wsSum}, the measurements indicate the prediction of the neutrino flux component of the anti-neutrino beam is consistent with a uniform reduction of $\sim$ 20\%.  These types of analyses may be of use to present and future precision neutrino experiments testing CP violation with neutrino and anti-neutrino beams in a non-magnetized environment.

%The method of fitting the muon angular distributions in the CCQE sample has already been employed in the MiniBooNE oscillation analysis~\cite{nubOsc1,nubOsc2}, while the CC1$\pip$- and $\mum$ capture-based measurements will likely be more valuable to anti-neutrino cross section extractions, as they are much less model-dependent and carry comparatively lower uncertainty.  

\section{Model Comparisons to $\numub$ CCQE Data}
\label{absModelComps}

With the largest background to the $\numub$ CCQE sample constrained by MiniBooNE's own data, meaningful measurements of $\numub$ CCQE can be made.  In this section the background-subtracted data is compared to the Fermi Gas model under various assumptions.  In the formalism, the so-called axial mass $M_{A}$ is the only free parameter for neutrino experiments to determine. Early measurements mostly on light nuclear targets provide a combined value of $M_{A} = 1.03 \pm 0.02$~GeV~\cite{MAMeas}.  The MiniBooNE measurement of $\numu$ CCQE on carbon yielded a significantly higher value of $M_{A} = 1.35 \pm 0.17$~GeV together with a mild scaling of nuclear Pauli blocking $\kappa = 1.007 \pm 0.012$~\cite{qePRD}.  Other high precision detectors with nuclear targets have also measured higher values of $M_{A}$ recently~\cite{K2KsciFi,MINOS,SciBooNE}.  
%The squared four-momentum transfer $Q^2_{QE}$ is particularly sensitive to $M_{A}$ and is given by

%\begin{equation}
%\label{eqn:QsqQE}
%Q^2_{QE} = 2 E_{\nub}^{QE} \left( p_\mu \textrm{cos}\,\thetmu - E_\mu \right) + m_\mu^2,
%\end{equation}

%\noindent where $E_{\nub}^{QE}$ is reconstructed from the muon information as in Equation~\ref{eqn:EnuPi} with the replacement of the neutron mass for the $\De$(1232) mass.

\begin{figure}[h]
\includegraphics[scale=0.46]{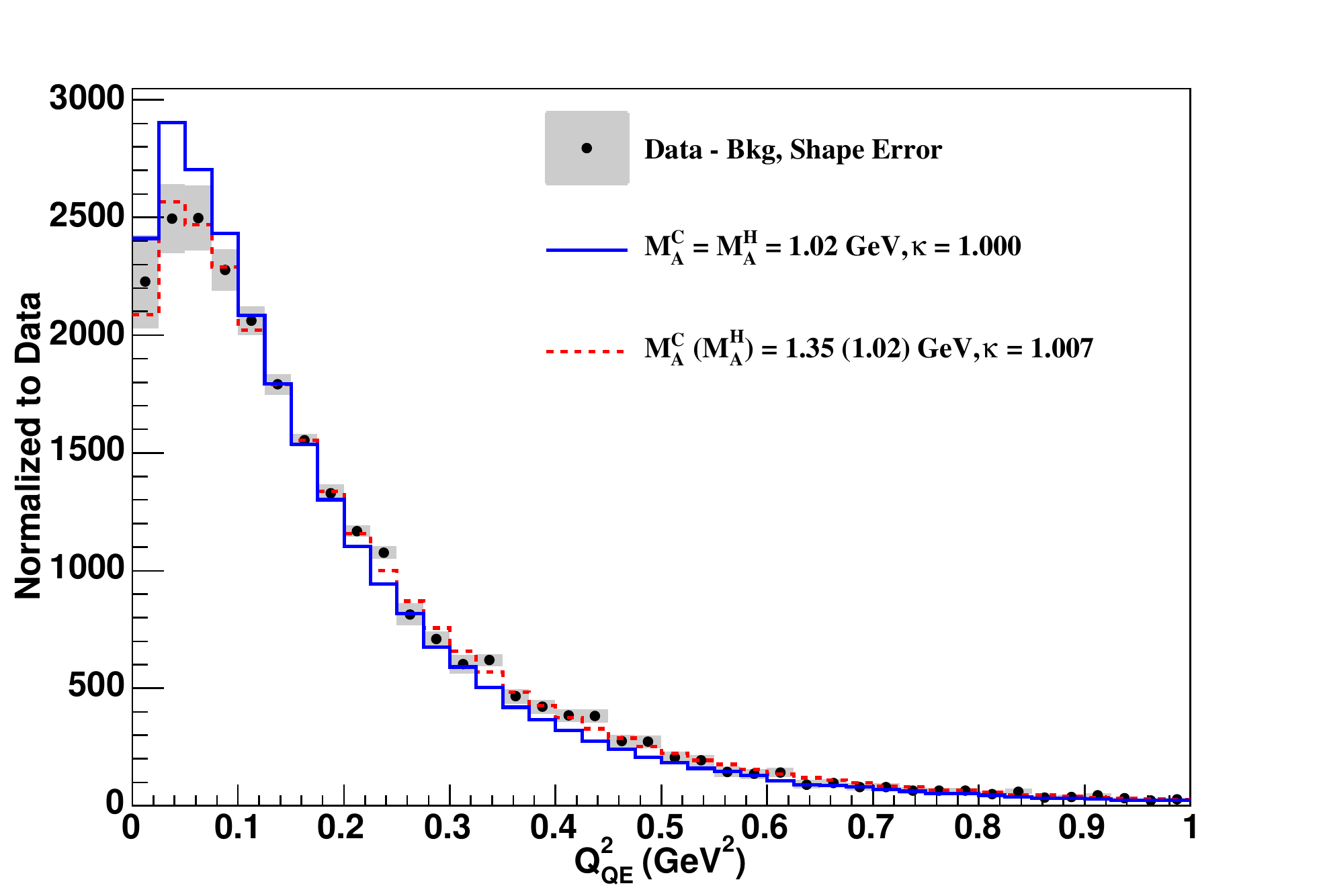} \\
\caption{(Color online) Shape comparisons of the squared four-momentum transfer, $Q^2_{QE}$, for MiniBooNE $\numub$ CCQE events and Fermi Gas model predictions with various $M_{A}$ assumptions. Neutrino-induced and non-QE backgrounds have been subtracted from the sample.}
\label{fig:compShape}
\end{figure}

\begin{figure}[h]
\includegraphics[scale=0.46]{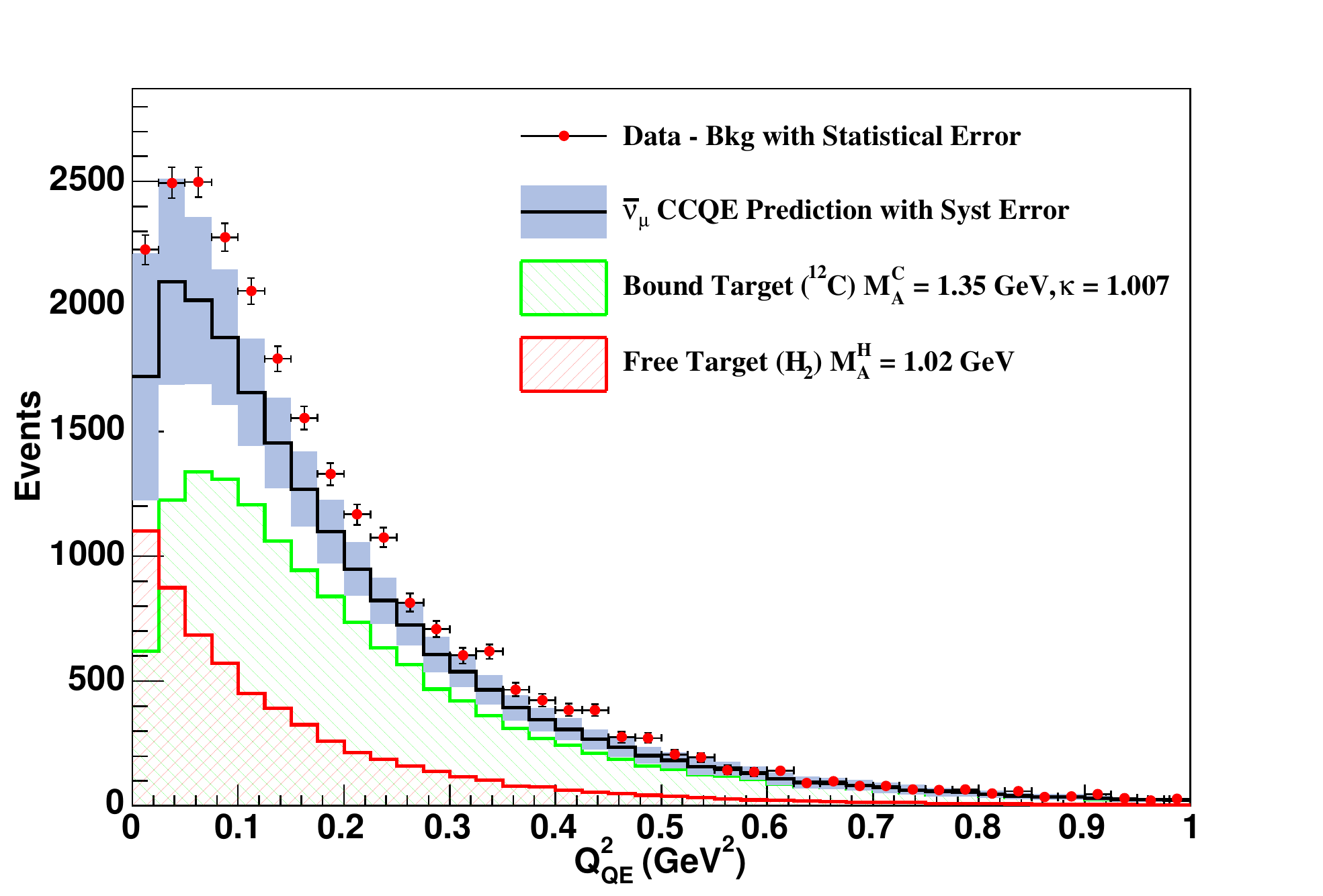} \\
\caption{(Color online) Absolutely-normalized comparison of $Q^2_{QE}$ for MiniBooNE $\numub$ CCQE events and the Fermi Gas prediction with $M_{A}^{C}$ ($M_A^H$) = 1.35 (1.02)~GeV.  The integrated ratio data/simulation for this choice is $1.21 \pm 0.12$.  Neutrino-induced and non-QE backgrounds have been subtracted from the sample.}
\label{fig:absComp}
\end{figure}

The MiniBooNE detector medium (C$_{n}$H$_{2n+2}$, $n\,\sim\,20$) contains a combination of quasi-free and bound proton targets for the $\numub$ CCQE interaction.  Using separate axial masses for the hydrogen (``$M_{A}^{H}$'') and carbon (``$M_{A}^{C}$'') scattering components, model comparisons to data are presented in Figures~\ref{fig:compShape} and~\ref{fig:absComp}.  For simultaneous consistency with the light target results and the neutrino-mode MiniBooNE measurement, the value of $M_{A}^{H}$ ($M_{A}^{C}$) is set to 1.02 (1.35)~GeV for the absolute comparison, while a shape comparison is made with this prediction and also with $M_{A}^{H}$ = $M_{A}^{C}$ = 1.02~GeV.  The latter choice does not describe the data shape, while the shape provided by the former parameter set is consistent but is also roughly 20\% low in normalization.

\section{Conclusion}
\label{conc}

The $\numu$ flux in the MiniBooNE $\numub$ beam has been measured with three independent and complementary techniques.  Results indicate the uncertain flux prediction is overestimated by $\sim$ 20\%, while the flux shape is well-modeled.  This measurement allows a much more precise evaluation of the $\numub$ CCQE data. The shape of these data are consistent with an axial mass value for bound nucleon targets of 1.35~GeV, while the ratio of the data to this prediction is 1.21 $\pm$ 0.12.   

\bibliographystyle{aipproc}

\end{document}